\documentclass[twocolumn,letterpaper,fleqn]{ModDetSymp}
\usepackage{fix2col}
\usepackage{multicol}
\usepackage{ifthen}
\usepackage{graphicx}
\usepackage{subfig}

{}
\newcommand{\AAA}{\AA\,}{}
{}

\newcommand{\Fig}[1]{Fig.~\ref{#1}}

\newcommand{\Tab}[1]{{Table~\ref{#1}}}

\title{A first-principles investigation of the
equation of states and molecular ``weak spots" 
of $\beta$-cyclotetramethylene tetranitramine (HMX)}

\author{Qing Peng and Suvranu De}

\affiliation{Department of Mechanical, Aerospace and Nuclear Engineering, \\
Rensselaer Polytechnic Institute, Troy, NY 12180, U.S.A.}

\begin{document}


\twocolumn[
\setlength{\fboxrule}{0.5pt}
\begin{@twocolumnfalse}
\maketitle
\begin{center}
\parbox{5in}{\textbf{Abstract. }%
We investigate the equation of states 
of the $\beta$-polymorph
of cyclotetramethylene tetranitramine (HMX)
energetic molecular crystal   
using DFT-D2,
a first-principles calculation based on
density functional theory (DFT) with van der Waals (vdW) corrections.
The atomic structures and equation of states under hydrostatic compressions
are studied for pressures up to 100 GPa. 
We found that the N-N bonds along the minor axis of the ring are 
more sensitive to the variation of pressure, 
which indicates that they are potential 
``weak spots" 
in atomic level within a single molecule of $\beta$-HMX.
Our study suggested that the van der Waals interactions 
are critically important in modeling this molecular crystal.


\sectionline}
\end{center}
\end{@twocolumnfalse}
]

\section{Introduction}

\begin{figure}[htp]
\subfloat[Unit cell]{\includegraphics[width=\linewidth]{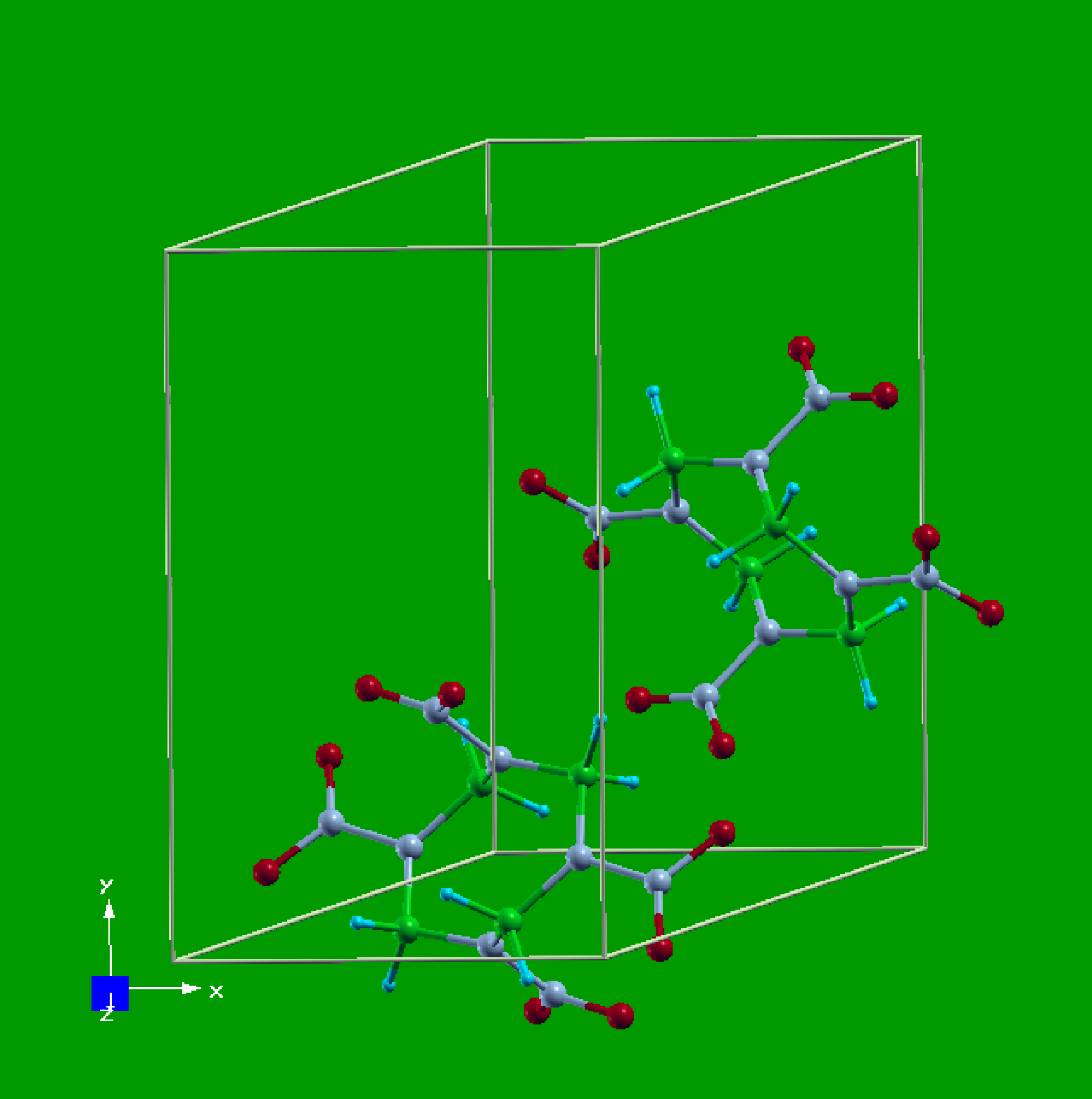}} \\
\subfloat[molecule]{\includegraphics[width=\linewidth]{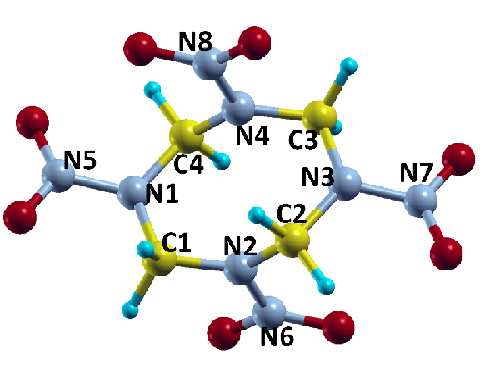}}
\caption
{\label{cell}
{\bf Geometry of $\beta$-HMX}
({\bf a}) The unit cell of $\beta$-HMX containing two HMX molecules with ring-chain structure.
The four carbon atoms (yellow) and four nitrogen atoms forms the ring-chain.
({\bf b}) A molecule with a ring-chain structure in the $\beta$-HMX crystal.
The molecule is center-symmetric.
The N1-N3 and N2-N4 forms major and minor axis of the ring-chain structure, respectively.
The four carbon atoms are coplanar on the C4-plane.
The four nitrogen atoms on the ring-chain are also coplanar on the N4-plane.
The angle between C4-plane and N4-plane is 30.3$^\circ$.
}
\end{figure}

Cyclotetramethylene-tetranitramine (HMX) is an important secondary explosive that is most commonly
used in polymer-bonded explosives (PBX), and as a solid rocket propellant. The $P2_1/c$ monoclinic 
$\beta$-phase molecular crystal (as shown in the \Fig{cell} (a))
is thermodynamically most stable polymorph of HMX at room temperature and
has highest density, which is an important factor in detonation velocity. 
A molecule with a ring-chain structure in the $\beta$-HMX crystal is shown in \Fig{cell} (b).
Due to its importance, HMX is subjected to extensive studies, both experimental 
\cite{Choi1970actaCrystal,doi:10.1063/1.480341,Herrmann1993,Deschamps2011,doi:10.1063/1.3211927} 
and theoretical \cite{Conroy2008PRB,Conroy2008JAPbetaHMX,doi:10.1021/jp809843k,Landerville2010APL}.
The anisotropic deformation in $\beta$-HMX molecular crystals determine how mechanical work 
is localized to form ``hot spots" that
promote rapid molecular decomposition, which is necessary for detonation 
\cite{AmirHMXjpd2010,De2014287}. 
These ``hot spots" are in general associated with the defects in the crystals, 
including voids, grain boundaries, and dislocations. 
In general, these defects could be used to engineer the materials' 
properties and performance \cite{rpi15,rpi8,Peng2010msmse7,rpi5,rpi21,Peng2010msmse10}. 
In atomistic models where there are no defects, 
the molecular crystals can sustain high pressures. 
Here we define the ``weak spots" 
as the region located at the weak bonds 
that are vulnerable to pressure or strains, 
which however are not well understood.

In addition, density plays an important role in the equation of state describing thermodynamic properties,
which can be used to describe materials at continuum level.
\cite{Landerville2010APL}
Despite its importance, the accurate prediction of the densities of energetic materials is challenging.
\cite{Huang2011EPL_hmx,rpi22}
Although first-principles calculations based on density functional theory (DFT)
provide overall better predictive power than force field models \cite{Peng2011jmps4},
they fail in predicting densities of energetic materials with standard approximations,
partially due to their poor descriptions of dispersion forces in molecular crystals.
\cite{Byrd2004,Byrd2007,Conroy2008JAPbetaHMX,Deschamps2011,doi:10.1063/1.4754130}

There are extensive studies to improve the modeling of dispersion interactions,
or van der Waals (vdW) interactions \cite{rpi16}. 
The DFT-D2 method \cite{Grimme2006} is a revised version of DFT-D1 method \cite{Grimme2004}
and has considerable improvement of its predecessor
with negligible extra computing demands comparing to the standard DFT calculations.
Its application in studying the structural properties and equation of state
of $\beta$-HMX under high pressure needs further investigation.

Here we model $\beta$-HMX using DFT-D2 method \cite{Grimme2006},
which describes the van der Waals interactions
as a simple pair-wise force field.
This method is chosen as a result of compromise between two opposite considerations:
accuracy and feasibility.
We investigate equation of state of the $\beta$-HMX and the atomic structures under high pressures. 
We found that the N-N bonds along the minor axis (N2-N6 and N4-N8 bonds in \Fig{cell} (b)) 
are weakest bonds thus atomic ``weak spots" in $\beta$-HMX molecules.

\section{FIRST-PRINCIPLES MODELLING}

We consider a conventional unit cell containing two HMX molecules (56 atoms in total)
with periodic boundary conditions, as depicted in \Fig{cell}).
The total energies of the system,
forces on each atom, stresses, and stress-strain relationships of $\beta$-HMX under the desired deformation configurations
are characterized via first-principles calculations based on density-functional theory (DFT).
DFT calculations were carried out with the Vienna Ab-initio
Simulation Package (VASP)
\cite{VASPa,VASPb,VASPc,VASPd} which is based
on the Kohn-Sham Density Functional Theory (KS-DFT) \cite{DFTa,DFTb} with the generalized gradient approximations
as parameterized by Perdew, Burke, and Ernzerhof (PBE)  for exchange-correlation functions \cite{PBE1,PBE2}.
The electrons explicitly included in the calculations are
the $1s^{1}$ for hydrogen atoms.
 $2s^{2}2p^4$ electrons for carbon atoms,
 $2s^{2}2p^5$ for nitrogen atoms,
and $2s^{2}2p^6$ for oxygen atoms.
The core electrons are replaced by the projector augmented wave (PAW) and pseudo-potential approach\cite{paw1,paw2}.
The kinetic-energy cutoff for the plane-wave basis was selected to be 800 eV in this study.
The calculations are performed at zero temperature.

The criterion to stop the relaxation of the electronic degrees of freedom is set by total energy change to be smaller than 0.000001 eV.
The optimized atomic geometry was achieved through minimizing Hellmann-Feynman forces
acting on each atom  until the maximum forces on the ions were smaller than 0.001 eV/\AA.
The atomic structures of all the deformed and undeformed configurations were obtained
by fully relaxing a 168-atom-unit cell.
The simulation invokes periodic boundary conditions for the three directions.
The irreducible Brillouin Zone was sampled with a $5\times3\times4$ Gamma-centered $k$-mesh.
The initial charge densities were taken as a superposition of atomic charge densities.

In the DFT-D2 method\cite{Grimme2006}, the van der Waals interactions are described using
a pair-wise force field. Such a semi-empirical dispersion potential is then added to the conventional
Kohn-Sham DFT energy as $E_{\mathrm{DFT-D2}}=E_{\mathrm{DFT}}+E_{\mathrm{disp}}$,
and
\begin{equation} 
\label{eq:shear}  
E_{\mathrm{disp}}=-\frac{1}{2} \sum_{i=1}^N \sum_{j=1}^N \sum_L' \frac{C_{6,ij}}{r^6_{ij,L}} f_{d,6}(r_ij,L), 
\end{equation}
where $N$ is the number of atoms.
The summations go over all atoms and all translations of the unit cell $L=(l_1,l_2,l_3)$.
The prime indicates that for $L=0$, $i\not=j$.
$C_{ij,6}$ stands for the dispersion coefficient for the atom pair $ij$.
$r_{ij,L}$ is the distance between atom $i$ in the reference cell $L=0$
and atom $j$ in the cell L.
$f(r)$ is a damping function whose role is to scale the force field
such as to minimize contributions from interactions within typical bonding distances $r$.
Since the van der Waals interactions decay quickly in the power of -6,
the contributions outside a certain suitably chosen cutoff radius are negligible.
The cutoff radius for pair interaction in this study is set to 30.0 \AA.
Here Fermi-type damping function is used as
\begin{equation} 
f_{d,6}(r_{ij})=\frac{s_6}{1+e^{-d(\frac{r_{ij}}{s_rR_{0ij}}-1)}}, 
\end{equation}
where $S_6$ is the global scaling parameter.
The global scaling factor $S_6=0.75$ is used for PBE  exchange-correlation functions.
$s_R$ is fixed at 1.00. The damping parameter $d=20.0$ is used.

\section{RESULTS AND ANALYSIS}

\subsection{\em Atomic structure and geometry}

\begin{table*}
\caption{\label{tab:a0} Lattice constants $a,b,c$, lattice angle $\beta$, volume of the unit cell $V$, and density $\rho$ predicted from DFT and DFT-D2 calculations, compared with experiments and previous calculations. The numbers in parentheses are differences in percentage referring to the experiment$^a$. }
\center
\begin{tabular}{c|c|c|c|c|c|c}
\hline
\hline
                 &$a$(\AAA)          & $b$(\AAA)        &$c$(\AAA)  & $\beta$ &V(\AAA$^3$)  & $\rho$ ($10^3Kg/m^3$) \\
\hline
Expt.$^a$  &6.54             &11.05     &8.70  &124.30  &519.39 &1.894\\
\hline
Expt.$^b$  &6.537            &11.054    &8.7018  &124.443  &518.558 &1.897 \\
\hline
Expt.$^c$  &6.5255           &11.0369   &7.3640  &102.670  &517.45&1.901 \\
\hline
Expt.$^d$  &6.54             &11.05     &7.37  &102.8  &519.37&1.894 \\
\hline
DFT     & 6.673(+2.0)   & 11.312(+2.4)    & 8.894(+2.2) &124.395(+0.1) &553.99(+6.7)&1.775(-6.2) \\  
\hline
{\bf DFT-D2}  & {\bf 6.542}(+0.0)   & {\bf 10.842}(-1.9)    &{\bf 8.745}(+0.5)  &{\bf 124.41}(+0.1)  &{\bf 511.73}(-1.5) &{\bf 1.923}(+1.5)     \\
\hline
Theory$^e$      & 6.70(+2.5)    &11.35(+2.7)      &8.91(+2.4)  &124.13(-0.1) &560.86(+8.0) &1.754(-7.4)        \\
\hline
Theory$^{f1}$   &6.38(-2.5)    &10.41(-5.8)       &8.43(-3.1)   &123.0(-1.1) &463.1(-10.7) &2.122(12.0)         \\
\hline
Theory$^{f2}$   &6.90(+5.5)    &11.65(+5.4)       &9.15(+5.2)  &124.5(+0.2) &608.1(+17.1)  &1.617(-14.6)       \\
\hline
Theory$^{f3}$   &6.56(+0.3)    &10.97(-0.7)       &8.70(0.0)    &124.4(+0.1) &517.4(-0.4)   &1.901(+0.4)        \\
\hline
Theory$^{g2}$   &6.78(+3.7)    &11.48(+3.9)       &9.19(+5.6)  &125.02(+0.6) &585.57(+12.7)&1.680(-11.3)       \\
\hline
Theory$^{g1}$   &6.43(-1.7)    &10.34(+6.4)       &8.61(-1.0)  &124.23(-0.1) &473.81(-8.8) &2.076(+9.6)        \\
\hline
Theory$^h$      &6.539(0.0)    &11.03(-0.2)      &8.689(-0.1) &123.9(-0.3) &520.17(+0.2)  &1.891(-0.2)        \\
\hline
Theory$^i$      &6.762(+3.4)    &11.461(+3.7)     &8.865(+1.9) &123.8(-0.4) &570.599(+9.9) &1.724(-9.0)        \\
\hline
Theory$^j$              &6.67(+2.0)    &11.17(+1.1)       &8.95(+2.9)  &124.5(+0.2) &549.30(+5.8)  &1.791(-5.5)        \\
\hline
Theory$^k$              &6.57(+0.5)    &11.02(-0.3)       &9.04(+3.9)  &124.9(+0.5) &531.12(+2.3)  &1.852(-2.2)        \\
\hline
Theory$^l$              &6.58(+0.6)    &11.12(+0.6)       &8.76(+0.7)  &124.3(+0.0) &529.8(+2.0)           &1.856(-2.0) \\
\hline
Theory$^m$              &6.57(+0.5)    &10.63(-3.8)       &9.13(+4.9)  &123.67(-0.5) &530.6(+2.2)   &1.854(-2.1)       \\
\hline
Theory$^{n1}$   &-   &-         &-      &- &556.07(+7.1) &1.769(-6.6)      \\
\hline
Theory$^{n2}$   &-   &-         &-      &- &500.77(-3.6) &1.964(+3.7)      \\
\hline
Theory$^{n3}$   &-   &-         &-      &- &519.41(+0.0) &1.894(0.0)        \\

\hline
\hline
\end{tabular}  \\

$^a$ Ref. \cite{Choi1970actaCrystal,doi:10.1063/1.480341}
$^{b}$ Ref. \cite{Herrmann1993}
$^{c}$ 303K space group $P2_1/n$ in Ref. \cite{Deschamps2011}
$^{d}$ space group $P2_1/n$ in Ref. \cite{doi:10.1063/1.3211927}
$^e$ DFT study using PAW-PBE(GGA) in Ref. \cite{Conroy2008JAPbetaHMX}
$^{f1}$ DFT study using USPP-LDA in Ref. \cite{doi:10.1063/1.3587135}
$^{f2}$ DFT study using USPP-PBE(GGA) in Ref. \cite{doi:10.1063/1.3587135}
$^{f3}$ DFT-D2 study using USPP-PBE(GGA) in Ref. \cite{doi:10.1063/1.3587135}
$^{g1}$ DFT study using USPP-LDA in Ref. \cite{Lu2008MP:HMX}
$^{g2}$ DFT study using USPP-PBE(GGA) in Ref. \cite{Lu2008MP:HMX}
$^{h}$ DFT study using USPP-LDA in Ref. \cite{Zhu2009TCA_DFT}
$^i$ DFT study using USPP-PBE(GGA) in Ref. \cite{doi:10.1021/jp9090969}
$^j$ Monte Carlo calculations in Ref. \cite{doi:10.1063/1.367168}
$^k$ Molecular Dynamics study in Ref. \cite{doi:10.1021/jp991202o}
$^l$ Molecular Dynamics study in Ref. \cite{doi:10.1021/je100009m}
$^m$ Molecular Dynamics study in Ref. \cite{Lu_MolecularPhysics2009}
$^{n1}$ DFT study using PAW-PBE(GGA) in Ref. \cite{Landerville2010APL}
$^{n2}$ DFT-D1 study using PAW-PBE(GGA) in Ref. \cite{Landerville2010APL}
$^{n3}$ DFT-D1 study using PAW-PBE(GGA) and zero-point energy corrections (T=300K) in Ref. \cite{Landerville2010APL}
\end{table*}

We first optimize the geometry of the monoclinic crystal (also shown in \Fig{cell})
by full relaxation of all the atoms and lattice constants.
The optimized lattice constants are: $a$ = 6.542 \AA, $b$ = 10.842 \AA, $c$ = 8.745 \AA,
$\alpha=\gamma$=90$^\circ$, and $\beta$= 124.413$^\circ$.
For the comparison, we also optimize the structure without the van der Waals corrections.
Our results of the lattice constants, the volume of the unit cell, and the densities are summarized in \Tab{tab:a0}
and compared to the experiment and previous theoretical results.
It shows that the standard DFT calculations give poor predictions.
For example, the volume of the unit cell is 6.7\% higher than the experimental value.
The prediction of the density of $\beta$-HMX from our DFT-D2 calculations are more accurate,
with a difference of -1.47\% compared to the experimental value.
This is a significant improvement over standard DFT calculations without van der Waals corrections.

In general, the molecular dynamics (MD) simulations predict better lattice vectors than the standard DFT calculations.
It is partially because the MD calculations include the van der Waals interactions.
Once the van der Waals corrections are introduced, the first-principles calculations
\cite{Landerville2010APL,doi:10.1063/1.3587135}
show good predictions with accuracy within 3\% of the experimental values.

\subsection{\em Equation of States}

\begin{figure}[htp]
\includegraphics[width=\linewidth]{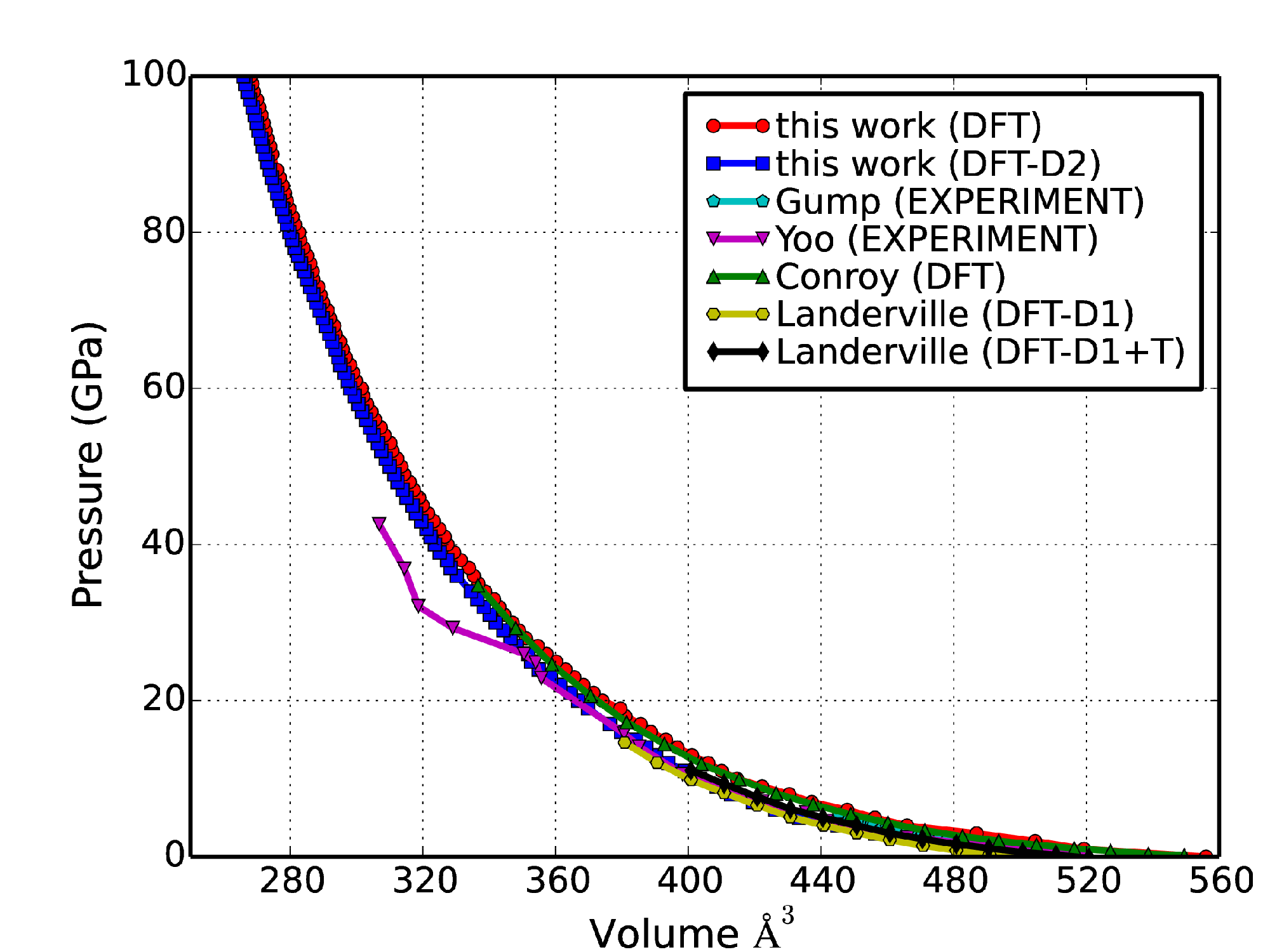} \\
\includegraphics[width=\linewidth]{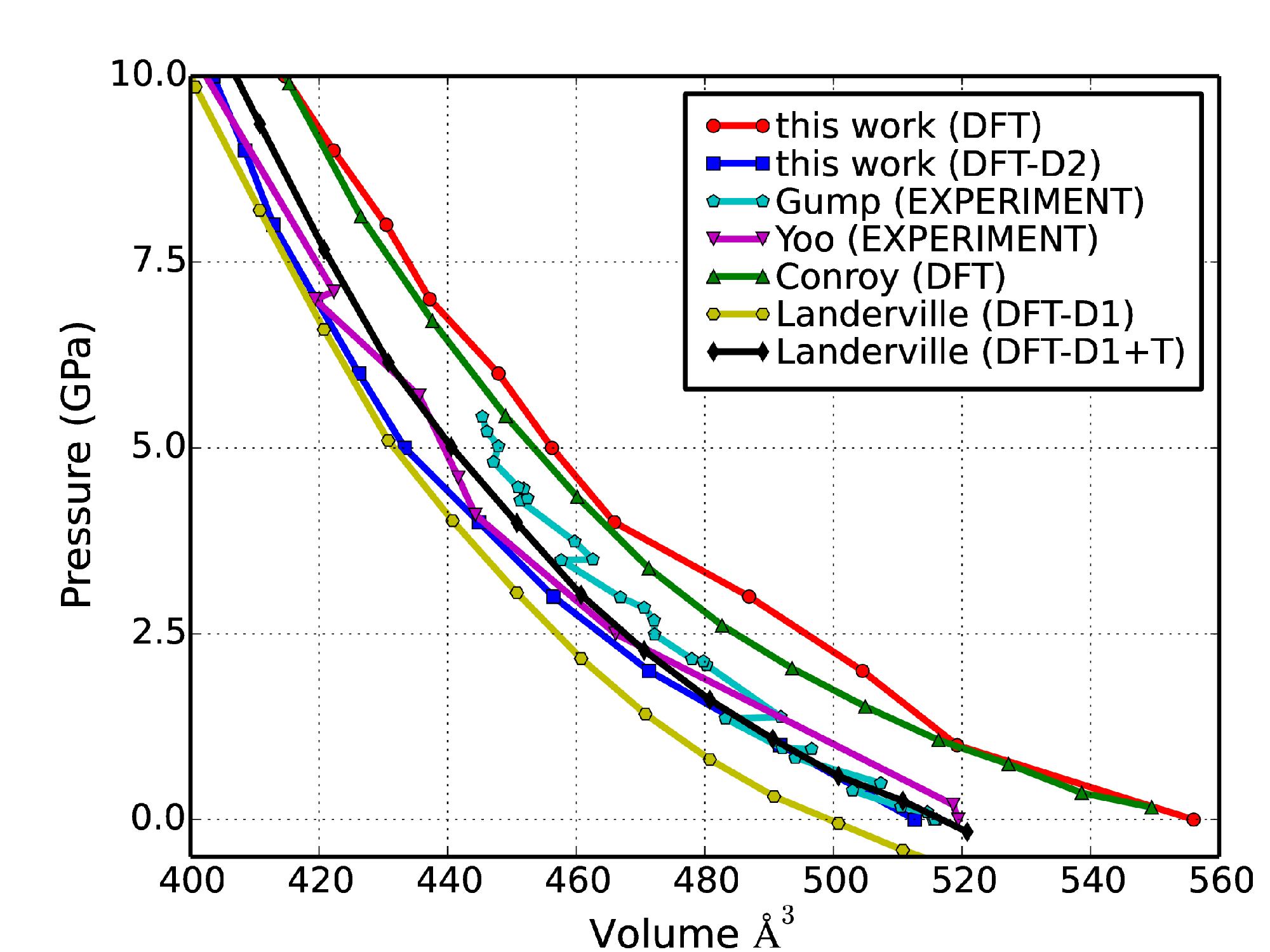}
\caption
{\label{pv}
{\bf Equations of States}
The Pressure-volume relationship of the solid $\beta$-HMX at zero temperature.
The volume is the 56-atom-unitcell volume.
The upper panel shows pressure from 0-100 GPa. The corresponding volume in the present DFT-D2 calculations
varies  from 512.64 $\AA^3$ (100\%) to 265.95 $\AA^3$ (51.9\%).
The lower panel shows pressure from 0-10 GPa for better comparison.
The experiments (Gump \cite{Gump2005JAP} and Yoo \cite{doi:10.1063/1.480341})
and previous calculations (Conroy \cite{Conroy2008JAPbetaHMX} and Landerville \cite{Landerville2010APL}).
}
\end{figure}

We study the isothermal equation of states of $\beta$-HMX
at zero temperature under hydrostatic pressures.
The corresponding volume is obtained after the system is fully relaxed.
The pressure-volume curve of unreacted $\beta$-HMX
at the temperature of 0K was illustrated in \Fig{pv}.
The volume corresponds to the 56-atom-unitcell volume.
The isothermal hydrostatic equation of state of $\beta$-HMX
is compared with experiments (Gump \cite{Gump2005JAP} and Yoo \cite{doi:10.1063/1.480341})
and previous calculations
(Conroy \cite{Conroy2008JAPbetaHMX} and Landerville \cite{Landerville2010APL})
.
The upper panel shows pressure from 0-100 GPa.
The corresponding volume in the present DFT-D2 calculations
varies  from 512.64 $\AA^3$ (100\%) to 265.95 $\AA^3$ (51.9\%).
The lower panel shows pressure from 0-10 GPa for better comparison.
Unlike standard DFT calculation (blue-circle line),
our DFT-D2 study (red-square line) shows reasonably good agreement with the hydrostatic-compression
experiments \cite{Gump2005JAP,doi:10.1063/1.480341}, 
suggesting that the Van der Waals interaction is critically important
in modeling the mechanical properties of this molecular crystal.

It is worthy pointing out that the experimental data were collected at room temperature.
Whereas, our results are for zero temperature
and we have not corrected the results to account for finite temperature.
Moreover, our calculations are performed using the $\beta$ polymorph of HMX, which is
consistent with the experimental data of Gump and Peiris. \cite{Gump2005JAP}
As observed in previous studies, \cite{Conroy2008PRB,doi:10.1063/1.2179801}
the calculated isotherm appears to approach experimental curve with increasing pressure.
However, as discussed below,
the sample from the experiment of Yoo and Cynn \cite{doi:10.1063/1.480341}
is no longer in the $\beta$ phase for
pressures beyond 12 GPa.

\subsection{\em ``Hot spots" within a molecule} 
\begin{figure}[htp]
\includegraphics[width=\linewidth]{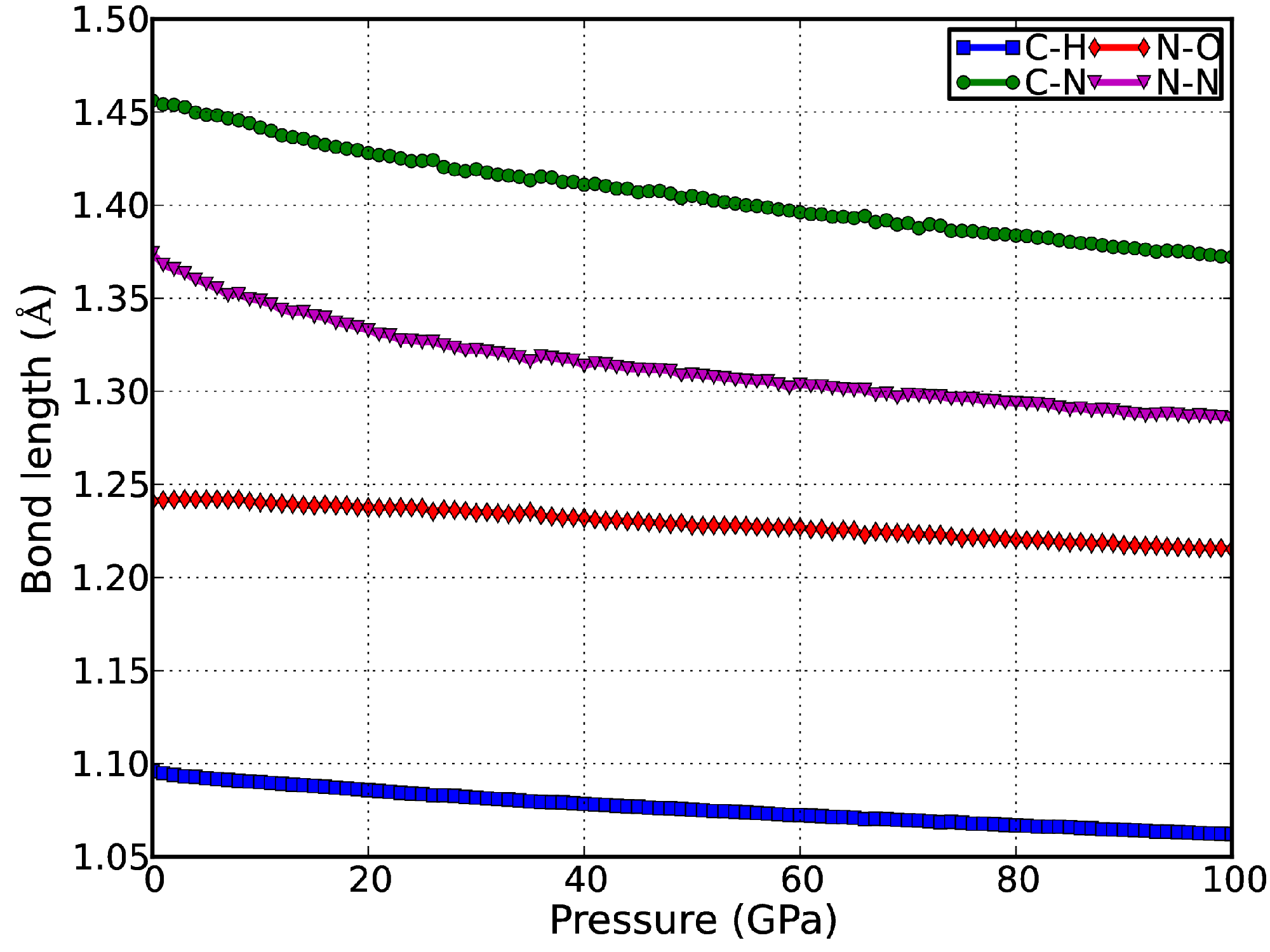} \\
\caption
{\label{bonds}
{\bf Bond lengths under high pressure}
The bond lengths of the bonds C-H, C-N, N-O, and N-N as a function of hydrostatic pressure $p$ ranged from 0-100 GPa.
The bond lengths are averaged over the unit cell.
}
\end{figure}

We next study the evolution of the structures of $\beta$-HMX under high pressures.
Firstly we studied the lattice structures including lattice constants and lattice angles.
We observed that the lattice constant $b$ and lattice angle $\beta$ are more sensitive to the applied pressures.
We then examine the bond lengths under various pressures in order to find the atomistic mechanism 
corresponding to the variations of pressure.
The bond lengths of the bonds C-H, C-N, N-O, and N-N as 
a function of hydrostatic pressure $p$ ranged from 0-100 GPa
are illustrated in \Fig{bonds}.
In our unit cell, the number of bonds is 16, 16, 16, and 8 for C-H, C-N, N-O, and N-N, respectively.
The bond lengths are averaged over the unit cell.

Our results show that the N-N bonds are the most sensitive to the applied hydrostatic pressure,
which indicates that they are potential atomistic ``weak spots" within a molecule 
because they are vulnerable to compression.

There are two kinds of N-N bonds in a HMX molecule:
one along the major axis along N1-N3 of the ring-chain (\Fig{cell}b)
and the other along the minor axis along N2-N4 of the ring-chain (\Fig{cell}b).
In order to find the more accurate atomistic ``weak spots" within a single molecule, 
we further examine the bending angles under various pressures.
The four carbon atoms are co-planar, marked as C4-plane.
Due to the symmetry, there are two angles between the N-N bonds and the C4-plane.
The angle along the major axis is denoted as $\beta_1$
and the angle along the minor axis is denoted as $\beta_2$.
During the compression, the two angles $\beta_1$ and $\beta_2$ might respond differently,
causing the anisotropy of the crystal.

\begin{figure}[htp]
\includegraphics[width=\linewidth]{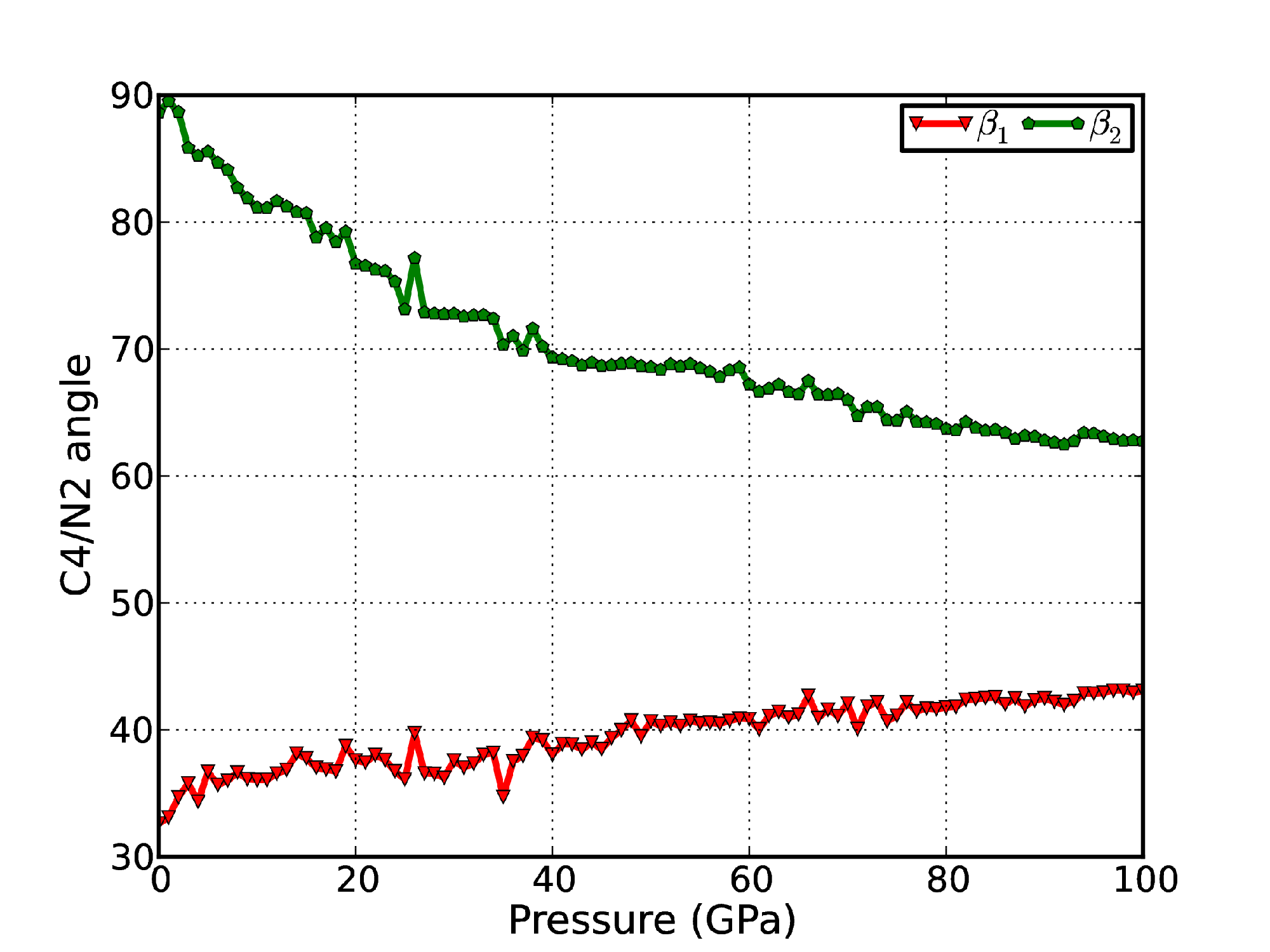} \\
\caption
{\label{NN}
{\bf Bending angles under high pressure}
The angle between the N-N bond and the plane formed by the four carbon atoms as a function of hydrostatic pressure $p$ ranged from 0-100 GPa.
}
\end{figure}

The $\beta_1$ and $\beta_2$ as a function of hydrostatic pressure $p$ ranged from 0-100 GPa were plotted in \Fig{NN}.
We found that $\beta_1$ increases with respect to an increasing pressure, opposite to the decrease of $\beta_2$.
Furthermore, the angle $\beta_2$ is more sensitive to the applied pressure than the angle $\beta_1$,
indicating that the N-N bonds along the minor axis are more vulnerable to the compression.
Therefore, we may conclude that the N-N bonds along the minor axis
is responsible for the sensitivity of $\beta$-HMX.

\section{CONCLUSIONS}

We studied the molecular structural and equation of states of
the $\beta$-HMX using the DFT-D2 calculations,
which is a first-principles calculation based on density
functional theory (DFT) with van der Waals corrections.
The accuracy of the density is significantly improved
from -6.2\% to +1.5\% using van der Waals corrections compared to the standard DFT calculations.
The equation of states under hydrostatic compressions are studied for pressures up to 100 GPa.
The agreement of our predictions with the experiments suggests that
the van der Waals interactions are critically important
in modeling the mechanical properties of
this molecular crystal.
Our analysis of the hydrostatic compression of $\beta$-HMX molecular structures
reveals that the lattice constant
$b$ are more sensitive to pressure.
We then studied the the bond lengths of C-H, C-N, N-O, and N-N bonds.
We found that the N-N bonds are vulnerable to the pressure.
We further studied the bending angle between N-N bonds with the plane formed by carbon atoms.
It turns out that the N-N bonds along the minor axis of the ring are more susceptible to pressure. 
Thus the ``weak spots" in atomic level are the N-N bonds along the minor axis of the ring 
within a single molecule of $\beta$-HMX crystals.

\section{ACKNOWLEDGEMENTS}
We thank
Michael W. Conroy,
Ramagopal Ananth,
Dongqing Wei,
and Shengbai Zhang
for helpful discussions.
The authors would like to acknowledge the generous financial support from
the Defense Threat Reduction Agency (DTRA) Grant \# BRBAA08-C-2-0130 and \# HDTRA1-13-1-0025.



\begin{thebibliography}{10}
\newcommand{\enquote}[1]{``#1''}

\bibitem{Choi1970actaCrystal}
Choi, C.~S. and Boutin, H.~P., \enquote{{A study of crystal structure of
  beta-cyclotetramethylene tetranitramine by neutron diffraction},} \emph{{Acta
  Crystallogr., Sect. B: Struct. Sci.}}, Vol. {B 26}, p. {1235}, {1970}.

\bibitem{doi:10.1063/1.480341}
Yoo, C.-S. and Cynn, H., \enquote{Equation of state, phase transition,
  decomposition of $\beta$-HMX
  (octahydro-1,3,5,7-tetranitro-1,3,5,7-tetrazocine) at high pressures,}
  \emph{J. Chem. Phys.}, Vol. 111, pp. 10229--10235, 1999.

\bibitem{Herrmann1993}
Herrmann, M., Engel, W. and Eisenreich, N., \enquote{Thermal-analysis Of The
  Phases Of Hmx Using X-ray Diffraction,} \emph{{Zeitschrift Fur
  Kristallographie}}, Vol. {204}, pp. 121--128, {1993}.

\bibitem{Deschamps2011}
Deschamps, J.~R., Frisch, M. and Parrish, D., \enquote{{Thermal Expansion of
  HMX},} \emph{{J. Chem. Crystallogr.}}, Vol.~{41}, pp. 966--970, {JUL} {2011}.

\bibitem{doi:10.1063/1.3211927}
Sun, B., Winey, J.~M., Gupta, Y.~M. and Hooks, D.~E., \enquote{Determination of
  second-order elastic constants of cyclotetramethylene tetranitramine
  ($\beta$-HMX) using impulsive stimulated thermal scattering,} \emph{J. Appl.
  Phys.}, Vol. 106, p. 053505, 2009.

\bibitem{Conroy2008PRB}
Conroy, M.~W., Oleynik, I.~I., Zybin, S.~V. and White, C.~T.,
  \enquote{First-principles investigation of anisotropic constitutive
  relationships in pentaerythritol tetranitrate,} \emph{Phys. Rev. B}, Vol.~77,
  p. 094107, Mar 2008.

\bibitem{Conroy2008JAPbetaHMX}
Conroy, M.~W., Oleynik, I.~I., Zybin, S.~V. and White, C.~T.,
  \enquote{{First-principles anisotropic constitutive relationships in
  beta-cyclotetramethylene tetranitramine (beta-HMX)},} \emph{{J. Appl.
  Phys.}}, Vol. {104}, p. {053506}, {SEP 1} {2008}.

\bibitem{doi:10.1021/jp809843k}
Conroy, M.~W., Oleynik, I.~I., Zybin, S.~V. and White, C.~T., \enquote{Density
  Functional Theory Calculations of Solid Nitromethane under Hydrostatic and
  Uniaxial Compressions with Empirical van der Waals Correction,} \emph{J.
  Phys. Chem. A}, Vol. 113, pp. 3610--3614, 2009.

\bibitem{Landerville2010APL}
Landerville, A.~C., Conroy, M.~W., Budzevich, M.~M., Lin, Y., White, C.~T. and
  Oleynik, I.~I., \enquote{{Equations of state for energetic materials from
  density functional theory with van der Waals, thermal, and zero-point energy
  corrections},} \emph{{Appl. Phys. Lett.}}, Vol.~{97}, p. {251908}, {DEC 20}
  {2010}.

\bibitem{AmirHMXjpd2010}
Zamiri, A.~R. and De, S., \enquote{{Deformation distribution maps of $beta$-HMX
  molecular crystals},} \emph{{J. Phys. D-Appl. Phys.}}, Vol.~{43}, p.
  {035404}, {JAN 27} {2010}.

\bibitem{De2014287}
De, S., Zamiri, A.~R. and Rahul, \enquote{A fully anisotropic single crystal
  model for high strain rate loading conditions with an application to
  $\alpha$-RDX,} \emph{J. Mech. Phys. Solids}, Vol.~64, pp. 287 -- 301, 2014.

\bibitem{rpi15}
Peng, Q., Crean, J., Dearden, A.~K., Wen, X., Huang, C., Bordas, S. P.~A. and
  De, S., \enquote{Defect engineering of 2D monatomic-layer materials,}
  \emph{Mod. Phys. Lett. B}, Vol.~27, p. 1330017, 2013.

\bibitem{rpi8}
Peng, Q., Ji, W. and De, S., \enquote{First-Principles study of the Effects of
  Mechanical Strains on the Radiation Hardness of Hexagonal Boron Nitride
  Monolayers,} \emph{{Nanoscale}}, Vol.~5, pp. 695--703, 2013.

\bibitem{Peng2010msmse7}
Peng, Q., Zhang, X. and Lu, G., \enquote{{Structure, mechanical and
  thermodynamic stability of vacancy clusters in Cu},} \emph{Model. Simul.
  Mater. Sci. Eng.}, Vol.~{18}, p. {055009}, {JUL} {2010}.

\bibitem{rpi5}
Peng, Q., Ji, W., Huang, H. and De, S., \enquote{Stability of
  Self-interstitials in hcp-{Zr},} \emph{{J. Nucl. Mater.}}, Vol. 429, pp.
  233--236, 2012.

\bibitem{rpi21}
Peng, Q., Ji, W., Lian, J., Chen, X.-J., Huang, H., Gao, F. and De, S.,
  \enquote{Pressure effect on stabilities of self-Interstitials in
  HCP-Zirconium,} \emph{{Sci. Rep.}}, Vol.~4, p. 5735, 2014.

\bibitem{Peng2010msmse10}
Peng, Q., Zhang, X., Huang, C., Carter, E.~A. and Lu, G., \enquote{{Quantum
  mechanical study of solid solution effects on dislocation nucleation during
  nanoindentation},} \emph{Model. Simul. Mater. Sci. Eng.}, Vol.~{18}, p.
  {075003}, {OCT} {2010}.

\bibitem{Huang2011EPL_hmx}
Song, H.~J. and Huang, F., \enquote{{Accurately predicting the structure,
  density, and hydrostatic compression of crystalline
  beta-1,3,5,7-tetranitro-1,3,5,7-tetraazacyclooctane based on its
  wave-function-based potential},} \emph{{Europhys. Lett.}}, Vol.~{95}, p.
  {53001}, {SEP} {2011}.

\bibitem{rpi22}
Peng, Q., Rahul, Wang, G., Liu, G.~R. and De, S., \enquote{Structures,
  Mechanical Properties, Equations of State, and Electronic Properties of
  $\beta$-HMX under Hydrostatic Pressures: A DFT-D2 study,} \emph{{Phys. Chem.
  Chem. Phys.}}, Vol.~16, pp. 19972--19983, 2014.

\bibitem{Peng2011jmps4}
Peng, Q. and Lu, G., \enquote{{A comparative study of fracture in Al: Quantum
  mechanical vs. empirical atomistic description},} \emph{J. Mech. Phys.
  Solids}, Vol.~{59}, pp. {775--786}, {APR} {2011}.

\bibitem{Byrd2004}
Byrd, E. F.~C., Scuseria, G.~E. and Chabalowski, C.~F., \enquote{{An ab initio
  study of solid nitromethane, HMX, RDX, and CL20: Successes and failures of
  DFT},} \emph{{J. Phys. Chem. B}}, Vol. {108}, pp. 13100--13106, {SEP 2}
  {2004}.

\bibitem{Byrd2007}
Byrd, E. F.~C. and Rice, B.~M., \enquote{{Ab initio study of compressed
  1,3,5,7-tetranitro-1,3,5,7-tetraazacyclooctane (HMX),
  cyclotrimethylenetrinitramine (RDX),
  2,4,6,8,10,12-hexanitrohexaazaisowurzitane (CL-20),
  2,4,6-trinitro-1,3,5-benzenetriamine (TATB), and pentaerythritol tetranitrate
  (PETN)},} \emph{{J. Phys. Chem. C}}, Vol. {111}, pp. 2787--2796, {FEB 15}
  {2007}.

\bibitem{doi:10.1063/1.4754130}
Klimes, J. and Michaelides, A., \enquote{Perspective: Advances and challenges
  in treating van der Waals dispersion forces in density functional theory,}
  \emph{J. Chem. Phys.}, Vol. 137, p. 120901, 2012.

\bibitem{rpi16}
Peng, Q., Chen, Z. and De, S., \enquote{A density functional theory study of
  the mechanical properties of graphane with van der Waals corrections,}
  \emph{{Mech. Adv. Mater. Struc.}}, p. DOI:10.1080/15376494.2013.839067, 2013.

\bibitem{Grimme2006}
Grimme, S., \enquote{{Semiempirical GGA-type density functional constructed
  with a long-range dispersion correction},} \emph{{J. Comput. Chem.}},
  Vol.~{27}, pp. 1787--1799, {NOV 30} {2006}.

\bibitem{Grimme2004}
Grimme, S., \enquote{{Accurate description of van der Waals complexes by
  density functional theory including empirical corrections},} \emph{{J.
  Comput. Chem.}}, Vol.~{25}, pp. 1463--1473, {SEP} {2004}.

\bibitem{VASPa}
Kresse, G. and Hafner, J., \enquote{Ab initio molecular dynamics for liquid
  metals,} \emph{Phys. Rev. B}, Vol.~47, p. 558, 1993.

\bibitem{VASPb}
Kresse, G. and Hafner, J., \enquote{Ab initio molecular dynamics simulation of
  the liquid-metal-amorphous-semiconductor transition in germanium,}
  \emph{Phys. Rev. B}, Vol.~49, p. 14251, 1994.

\bibitem{VASPc}
Kresse, G. and Furthuller, J., \enquote{Efficient iterative schemes for ab
  initio total-energy calculations using a plane-wave basis set,} \emph{Phys.
  Rev. B}, Vol.~54, p. 11169, 1996.

\bibitem{VASPd}
Kresse, G. and Furthuller, J., \enquote{Efficiency of ab-initio total energy
  calculations for metals and semiconductors using a plane-wave basis set,}
  \emph{Comput. Mater. Sci.}, Vol.~6, p.~15, 1996.

\bibitem{DFTa}
Hohenberg, P. and Kohn, W., \enquote{Inhomogeneous Electron Gas,} \emph{Phys.
  Rev.}, Vol. 136, p. B864, Nov 1964.

\bibitem{DFTb}
Kohn, W. and Sham, L.~J., \enquote{Self-Consistent Equations Including Exchange
  and Correlation Effects,} \emph{Phys. Rev.}, Vol. 140, p. A1133, Nov 1965.

\bibitem{PBE1}
Perdew, J.~P., Burke, K. and Ernzerhof, M., \enquote{Generalized Gradient
  Approximation Made Simple,} \emph{Phys. Rev. Lett.}, Vol.~77, p. 3865, 1996.

\bibitem{PBE2}
Perdew, J.~P., Burke, K. and Ernzerhof, M., \enquote{Erratum: Generalized
  Gradient Approximation Made Simple,} \emph{Phys. Rev. Lett.}, Vol.~78, p.
  1396, 1997.

\bibitem{paw1}
Bl\"ochl, P.~E., \enquote{Projector augmented-wave method,} \emph{Phys. Rev.
  B}, Vol.~50, pp. 17953--17979, Dec 1994.

\bibitem{paw2}
Jones, R.~O. and Gunnarsson, O., \enquote{The density functional formalism, its
  applications and prospects,} \emph{Rev. Mod. Phys.}, Vol.~61, pp. 689--746,
  Jul 1989.

\bibitem{doi:10.1063/1.3587135}
Wu, Z., Kalia, R.~K., Nakano, A. and Vashishta, P., \enquote{Vibrational and
  thermodynamic properties of $\beta$-HMX: A first-principles investigation,}
  \emph{J. Chem. Phys.}, Vol. 134, p. 204509, 2011.

\bibitem{Lu2008MP:HMX}
Lu, L.-Y., Wei, D.-Q., Chen, X.-R., Lian, D., Ji, G.-F., Zhang, Q.-M. and Gong,
  Z.-Z., \enquote{{The first principle studies of the structural and
  vibrational properties of solid beta-HMX under compression},} \emph{{Mol.
  Phys.}}, Vol. {106}, pp. 2569--2580, {2008}.

\bibitem{Zhu2009TCA_DFT}
Zhu, W., Zhang, X., Wei, T. and Xiao, H., \enquote{DFT studies of pressure
  effects on structural and vibrational properties of crystalline
  octahydro-1,3,5,7-tetranitro-1,3,5,7-tetrazocine,} \emph{Theor. Chem. Acc.},
  Vol. 124, pp. 179--186, 2009.

\bibitem{doi:10.1021/jp9090969}
Cui, H.-L., Ji, G.-F., Chen, X.-R., Zhu, W.-H., Zhao, F., Wen, Y. and Wei,
  D.-Q., \enquote{{First Principles Study of High Pressure Behavior of Solid
  beta HMX},} \emph{{J. Phys. Chem. A}}, Vol. 114, pp. 1082--1092, 2010.

\bibitem{doi:10.1063/1.367168}
Sewell, T.~D., \enquote{Monte Carlo calculations of the hydrostatic compression
  of hexahydro-1,3,5-trinitro-1,3,5-triazine and
  beta-octahydro-1,3,5,7-tetranitro-1,3,5,7-tetrazocine,} \emph{{J. Appl.
  Phys.}}, Vol.~{83}, pp. 4142--4145, {APR 15} {1998}.

\bibitem{doi:10.1021/jp991202o}
Sorescu, D.~C., Rice, B.~M. and Thompson, D.~L., \enquote{Theoretical Studies
  of the Hydrostatic Compression of RDX, HMX, HNIW, and PETN Crystals,}
  \emph{{J. Phys. Chem. B}}, Vol. 103, pp. 6783--6790, 1999.

\bibitem{doi:10.1021/je100009m}
Cui, H.-L., Ji, G.-F., Chen, X.-R., Zhang, Q.-M., Wei, D.-Q. and Zhao, F.,
  \enquote{Phase Transitions and Mechanical Properties of
  Octahydro-1,3,5,7-tetranitro-1,3,5,7-tetrazocine in Different Crystal Phases
  by Molecular Dynamics Simulation,} \emph{{J. Chem. Eng. Data}}, Vol.~55, pp.
  3121--3129, 2010.

\bibitem{Lu_MolecularPhysics2009}
Lu, L.-Y., Wei, D.-Q., Chen, X.-R., Ji, G.-F., Wang, X.-J., Chang, J., Zhang,
  Q.-M. and Gong, Z.-z., \enquote{{The pressure-induced phase transition of the
  solid beta-HMX},} \emph{{Mol. Phys.}}, Vol. {107}, pp. 2373--2385, {2009}.

\bibitem{Gump2005JAP}
Gump, J.~C. and Peiris, S.~M., \enquote{{Isothermal equations of state of beta
  octahydro-1,3,5,7-tetranitro-1,3,5,7-tetrazocine at high temperatures},}
  \emph{{J. Appl. Phys.}}, Vol.~{97}, p. {053513}, {MAR 1} {2005}.

\bibitem{doi:10.1063/1.2179801}
Liu, H., Zhao, J., Wei, D. and Gong, Z., \enquote{Structural and vibrational
  properties of solid nitromethane under high pressure by density functional
  theory,} \emph{J. Chem. Phys.}, Vol. 124, 124501, 2006.

\end{thebibliography}

\end{document}